\documentclass[12pt,aps,prd,preprint,onecolumn]{revtex4}
\usepackage{amssymb}
\usepackage{amsmath}
\usepackage{graphics}
\usepackage{latexsym}
\usepackage{array}
\usepackage[activeacute,english]{babel}
\usepackage[latin1]{inputenc}
\usepackage[dvips]{graphicx}
\usepackage{color}
\usepackage{slashed}

\newcommand{\bn}{\begin{eqnarray}}
\newcommand{\en}{\end{eqnarray}}
\newcommand{\be}{\begin{equation}}
\newcommand{\ee}{\end{equation}}

\newcommand{\bc}{\begin{center}}
\newcommand{\ec}{\end{center}}

\newcommand{\ket}[1]{\ensuremath{\left|#1\right\rangle}}
\newcommand{\bra}[1]{\ensuremath{\left\langle#1\right|}}

\newcommand{\dder}[2]{\frac{d^{2}#1}{d#2^{2}}}
\newcommand{\der}[2]{\frac{d#1}{d#2}}
\newcommand{\parc}[2]{\frac{\partial#1}{\partial #2}}

\newcommand{\abs}[1]{\left\vert#1\right\vert}
\newcommand{\crea}[2]{\hat{#1}^{\dag}_{#2}}
\newcommand{\des}[2]{\hat{#1}_{#2}}
\newcommand{\esp}[1]{\left[#1\right]}
\newcommand{\corc}[1]{\left(#1\right)}
\newcommand{\llav}[1]{{\left\{#1\right\}}}
\newcommand{\Dirac}{(i\gamma^{\mu}\partial_{\mu}-m)}

\newcommand{\DiracOsc}{(i\gamma^{\mu}\partial_{\mu}-m+\sigma^{\mu\nu}F_{\mu\nu})}
\begin{document}

\title{\Large \bf Canonical quantization of the Dirac oscillator field in (1+1) and (3+1) dimensions}

\author{C. J. Quimbay\footnote{Associate researcher of Centro
Internacional de F\'{\i}sica, Bogot\'a D.C., Colombia.}}
\email{cjquimbayh@unal.edu.co}

\affiliation{Departamento de F\'{\i}sica, Universidad Nacional de Colombia.\\
Ciudad Universitaria, Bogot\'{a} D.C., Colombia.}

\author{Y. F. P\'erez}
\email{yuber.perez@uptc.edu.co}

\affiliation{Escuela de F\'isica, Universidad Pedag\'ogica y
Tecnol\'ogica de Colombia, Tunja, Colombia}

\author{R. A. Hernandez}
\email{rahernandezm@unal.edu.co}

\affiliation{Departamento de F\'{\i}sica, Universidad Nacional de Colombia.\\
Ciudad Universitaria, Bogot\'{a} D.C., Colombia.}

\date{\today}

\begin{abstract}
The main goal of this work is to study the Dirac oscillator as a
quantum field using the canonical formalism of quantum field
theory and to develop the canonical quantization procedure for
this system in $(1+1)$ and $(3+1)$ dimensions. This is possible
because the Dirac oscillator is characterized by the absence of
the Klein paradox and by the completeness of its eigenfunctions.
We show that the Dirac oscillator field can be seen as constituted
by infinite degrees of freedom which are identified as decoupled
quantum linear harmonic oscillators. We observe that while for the
free Dirac field the energy quanta of the infinite harmonic
oscillators are the relativistic energies of free particles, for
the Dirac oscillator field the quanta are the energies of
relativistic linear harmonic oscillators.\\

\vspace{0.4cm} \noindent {\it{Keywords:}} {Quantum field theory,
Dirac oscillator, canonical quantization formalism, relativistic
harmonic oscillators.}

\vspace{0.4cm} \noindent {\it{PACS Codes:}}  11.10.Wx, 11.15.Ex,
14.70.Fm.

\end{abstract}

\maketitle


\section{Introduction}\label{sec:01}

Quantum states of a relativistic massive fermion are described by
four-components wave functions called Dirac spinors. These wave
functions, which are solutions of the Dirac equation, describe
states of positive and negative energy. For the case where
fermions carry the electric charge $q$, the electromagnetic
interaction of fermions can be included by means of the
electromagnetic fourpotential $A^\mu$, which is introduced in the
Dirac equation through the so called minimal substitution,
changing the fourmomentum such as $p^\mu \rightarrow p^\mu - q
A^\mu$. It is also possible to introduce a linear harmonic
potential in the Dirac equation by substituting $\vec{p}
\rightarrow \vec{p} - im\omega\beta \vec{r}$, where $m$ is the
fermion mass, $\omega$ represents an oscillator frequency, $r$ is
the distance of the fermion respects to the origin of the linear
potential and $\beta=\gamma^0$ corresponds to the diagonal Dirac
matrix.

The Dirac equation including the linear harmonic potential was
initially studied by It\^o et al. \cite{DI67},  Cook \cite{CC71}
and Ui et al. \cite{UI84}. This system was latterly called by
Moshinsky and Szczepaniak as Dirac oscillator \cite{MS89}, because
it behaves as an harmonic oscillator with a strong spin-orbit
coupling in the non-relativistic limit. As a relativistic quantum
mechanical system, the Dirac oscillator has been widely studied.
Several properties from this system have been considered in (1+1),
(2+1), (3+1) dimensions \cite{MS96}-\cite{YL11}. Specifically, for
the Dirac oscillator have been studied several properties as its
covariance \cite{MM89}, its energy spectrum, its corresponding
eigenfunctions and the form of the electromagnetic potential
associated with its interaction in (3+1) dimensions \cite{JB90},
its Lie Algebra symmetries \cite{CQ90}, the conditions for the
existence of bound states \cite{DA90}, its connection with
supersymmetric (non-relativistic) quantum mechanics \cite{BD90},
the absence of the Klein paradox in this system \cite{DA92}, its
conformal invariance \cite{RM92}, its complete energy spectrum and
its corresponding eigenfunctions in (2+1) dimensions \cite{VV94},
the existence of a physical picture for its interaction
\cite{RM95}. For this system, other aspects have been also studied
as the completeness of its eigenfunctions in (1+1) and (3+1)
dimensions \cite{RS01}, its thermodynamic properties in (1+1)
dimensions \cite{PL03}, the characteristics of its two-point Green
functions \cite{AA04}, its energy spectrum in the presence of the
Aharonov-Bohm effect \cite{FB04}, the momenta representation of
its exact solutions \cite{CQ05}, the Lorenz deformed covariant
algebra for the Dirac oscillator in (1+1) dimension \cite{CQ06},
the properties of its propagator in (1+1) dimensions using the
supersymmetric path integral formalism \cite{RR07}, its exact
mapping onto a Jaynes-Cummings model \cite{AB07}, its
nonrelativistic limit in (2+1) dimensions interpreted in terms of
a Ramsey-interferometry effect \cite{AB081}, the existence of a
chiral phase transition for this system in (2+1) dimensions in
presence of a constant magnetic field \cite{AB082}, a new
representation for its solutions using the Clifford algebra
\cite{RL08}, its dynamics in presence of a two-component external
field \cite{ES10}, the relativistic Landau levels for this system
in presence of a external magnetic field in (2+1) dimensions
\cite{MV10} and its relationship with (Anti)-Jaynes-Cummings
models in a (2+1) dimensional noncommutative space \cite{YL11}.

Some possible applications of the Dirac oscillator have been
developed. For instance, the hadronic spectrum has been studied
using the two-body Dirac oscillator in \cite{MS93,MS96} and
the references therein. The Dirac oscillator in (2+1) dimension
has been used as a framework to study some condensed matter
physical phenomena such as the study of electrons in two dimensional
materials, which can be applied to study some aspects of the physics
of graphene \cite{ES11}. This system has also been used in quantum
optics to describe the interaction of atoms with electromagnetic fields
in cavities (the Jaynes-Cummings model) \cite{ES11}.

The standard point of view of the Quantum Field Theory (QFT)
establishes that an excitation of one of the infinite degrees of
freedom that constitute the free Dirac quantum field can be
interpreted as a free relativistic massive fermion \cite{GR96}.
The free Dirac quantum field which describes the quantum dynamics
of a non-interacting relativistic massive fermion can be seen as
constituted by infinite decoupled quantum harmonic oscillators. In
this scheme a non-interacting relativistic massive fermion is
described as an excitation of a degree of freedom of the free
Dirac quantum field \cite{GR96}. For the free Dirac field the
energy quanta of the infinite harmonic oscillators are
relativistic energies of free fermions or antifermions. An
analogous situation is presented in the description of a
non-interacting relativistic boson as an excitation of one of the
infinite degrees of freedom (harmonic oscillators) that constitute
a free bosonic quantum field \cite{GR96}. For the free Boson field
the energy quanta of the infinite harmonic oscillators are
relativistic energies of free bosons.

On the other hand, the Dirac oscillator, which describes the
interacting system constituted by a relativistic massive fermion
under the action of a linear harmonic potential, has not been
studied as a quantum field up to now. In this direction, we show
that it is possible to study consistently the Dirac oscillator as
a QFT system. The main goal of this work is to consider the Dirac
oscillator as a field and perform the canonical quantization
procedure of the Dirac oscillator field in (1+1) and (3+1)
dimensions. We can do it, in the first place, because the Dirac
oscillator is characterized by the absence of the Klein paradox
\cite{DA92}, that allows to distinguish between positive and
negative energy states \cite{BL07}. In the second place, the Dirac
oscillator is an interacting system in which there exist bound
states that are defined by means of specific quantum numbers. This
aspect means that the solution of the Dirac oscillator equation
leads to both a well-known energy spectrum and an eigenfunction
set which satisfies the completeness and orthonormality conditions
in (1+1) and (3+1) dimensions \cite{RS01}. Thus the Dirac
oscillator field can be consistently treated as a quantum field
because it can be written in terms of a Fourier expansion of that
eigenfunction set. We observe that after the canonical
quantization procedure, the Dirac oscillator field can be seen as
constituted by infinite relativistic decoupled quantum linear
harmonic oscillators.

The canonical quantization procedure for the Dirac oscillator
field starts from writing the Hamiltonian operator for this system
in terms of annihilation and creation operators satisfying the
usual anticommutation relations. This procedure allow us to obtain
the Feynman propagator for the Dirac oscillator field. Thus we are
able to study the differences between the Dirac oscillator field
and the free Dirac field. We note that while for the free Dirac
field the energy quanta of the infinite harmonic oscillators are
relativistic energies of free particles, for the Dirac oscillator
quantum field the quanta are energies of relativistic linear
harmonic oscillators.

This work is presented as follows: First, in section 2 we
introduce the notation used in this paper presenting the relevant
aspects of the standard canonical quantization procedure for the
free Dirac field in 3+1 dimensions. Next, in section 3 we study
the Dirac oscillator system in 1+1 dimensions from a point of view
of quantum relativistic mechanics; additionally we describe some
aspects of the Dirac's sea picture for this system and we develop
the canonical quantization procedure for the Dirac oscillator
field in 1+1 dimensions. Posteriorly, in section 4 we study the
canonical quantization procedure for the Dirac oscillator field in
3+1 dimensions. Finally, in section 5 we present some conclusions
of this work.


\section{Canonical quantization for the free Dirac field}\label{sec:CuantDirLib}

In this section we present the main steps of the standard
canonical quantization procedure of the free Dirac field following
the procedure developed in \cite{GR96}. The equation of motion for
a free relativistic fermion of mass $m$ is given by the Dirac
equation
\begin{equation}
\Dirac\psi(\vec r,t)=0,
\end{equation}
where we have taken $\hbar=c=1$. In this equation $\gamma^\mu$
represents the Dirac matrices obeying the Clifford algebra
$\{\gamma^\mu,\gamma^\nu\}=2g^{\mu\nu}$ and $\psi(\vec r,t)$
represents the four-component spinor wave functions. The four
linearly independent solutions from the Dirac equation are plane
waves of the form $\psi_{\vec p}^r(\vec
r,t)=(2\pi)^{-3/2}\sqrt{\frac{m}{E_{\vec p}}}w_r(\vec
p)e^{-i\epsilon_r(E_{\vec p} t-\vec{p}\cdot\vec{r})}$, where
$E_{\vec p}$ is given by $E_{\vec p}=+\sqrt{\vec p^{\ 2}+m^2}$.
The two solutions of relativistic free-particle positive energy
are described by $r=1,2$ while the two solutions of relativistic
free-particle negative energy are described by $r=3,4$. The sign
function $\epsilon_r$ takes the value $\epsilon_r=1$, for $r=1,2$,
and $\epsilon_r=-1$, for $r=3,4$. Additionally, the spinors
$w_r(\vec p)$ obey the equation $(\gamma_\mu p^\mu-\epsilon_r
m)w_r(\vec p)=0$.

In a quantum field theory treatment, $\psi(\vec r,t)$ is the free
Dirac field. The Hamiltonian associated to this field is $
H=\int\,d^3x\psi^\dag(-i\boldsymbol{\alpha}\cdot\nabla+m\beta)\psi$,
where the matrices $\boldsymbol{\alpha},\beta$ are defined as
$\alpha_i=\gamma^0\gamma^i$, $\beta=\gamma^0$. The canonical
quantization procedure of the free Dirac field starts by replacing
the fields $\psi(\vec{r},t)$ and $\psi^\dag(\vec{r},t)$ with the
fields operators $\hat\psi(\vec{r},t)$ and
$\hat\psi^\dag(\vec{r},t)$ which obey the usual commutation
relations of the Jordan-Wigner type \cite{GR96}. The Fourier
expansion of the free Dirac field operator $\hat\psi(\vec{r},t)$,
in terms of the plane waves functions, is written as
\begin{align}
\hat\psi(\vec{r},t)=\sum_{r=1}^4\int\frac{d^3p}{(2\pi)^{3/2}}
\sqrt{\frac{m}{E_{\vec p}}}\hat{a}(\vec{p},r)w_r(\vec
p)e^{-i\epsilon_rp\cdot x},
\end{align}
where $p\cdot x\equiv p_\mu x^\mu=E_{\vec p}
t-\vec{p}\cdot\vec{r}$ and $\hat{a}(\vec{p},r)$ is an operator.
The operators $\hat{a}(\vec{p},r)$ and its conjugated
$\hat{a}^\dag(\vec{p},r)$ satisfy the usual anticommutation
relations. The Hamiltonian operator $\hat{H}$ associated to this
system is defined in terms of the Dirac field operators
$\hat\psi(\vec{r},t)$ and $\hat\psi^\dag(\vec{r},t)$. Using the
Fourier expansion of the free Dirac field operator, $\hat{H}$ can
be written as $\hat{H}=\int d^3p E_{\vec p} \esp{ \sum_{r=1}^2
{\hat n}_{{\vec p},{\vec r}} +\sum_{r=3}^4 {\hat {\bar n}}_{{\vec
p},{\vec r}}}$, where ${\hat n}_{{\vec p},{\vec
r}}=\hat{a}^\dag(\vec{p},r)\hat{a}(\vec{p},r)$ is the
particle-number operator and ${\hat {\bar n}}_{{\vec p},{\vec
r}}=\hat{a}(\vec{p},r)\hat{a}^\dag(\vec{p},r)$ is the
antiparticle-number operator. In this Hamiltonian operator, which
is positively defined, the Dirac's see picture has been used and
the zero-point energy contribution has been subtracted
\cite{GR96}.

Now it is possible to introduced the following canonical
transformations over the annihilation and creation operators that
allow to differentiate between operators associated to particles
and antiparticles: $ \hat c(\vec p,+s)=\hat a(\vec p,1)$, $\hat
c(\vec p,-s)=\hat a(\vec p,2)$, $\hat d^\dag(\vec p,+s)=\hat
a(\vec p,3)$, $\hat d^\dag(\vec p,-s)=\hat a(\vec p,4)$. With
these transformations, the following identifications are possible:
$\hat c(\vec p,s)$ and $\hat c^\dag(\vec p,s)$ represent the
annihilation and the creation operators of particles,
respectively; $\hat d(\vec p,s)$ and $\hat d^\dag(\vec p,s)$
represent the annihilation and the creation operators of
antiparticles, respectively. Using the operators $\hat c(\vec
p,s)$, $\hat c^\dag(\vec p,s)$, $\hat d(\vec p,s)$ and $\hat
d^\dag(\vec p,s)$, the Hamiltonian operator of the free Dirac
field can be written as \cite{GR96}
\begin{align}
\hat{H}=\sum_{s}\int d^3p\,E_{\vec p}\corc{{\hat n}^{(c)}_{{\vec
p},s}+{\hat n}^{(d)}_{{\vec p},s}},\label{eq:HOFDF}
\end{align}
where ${\hat n}^{(c)}_{{\vec
p},s}=\hat{c}^\dag(\vec{p},s)\hat{c}(\vec{p},s)$ represents the
fermion-number operator and ${\hat n}^{(d)}_{{\vec
p},s}=\hat{d}^\dag(\vec{p},s)\hat{d}(\vec{p},s)$ represents the
antifermion-number operator. It is possible to observe that
$\hat{c}^\dag(\vec{p},s)$ creates a fermion of mass $m$ having
energy $E_{\vec p}$, momentum $\vec{p}$, charge $+e$ and
projection of spin $+s$, whereas $\hat{d}^\dag(\vec{p},s)$ creates
an antifermion of the same mass having the same energy, the same
momentum, charge $-e$ and projection of spin $-s$. In this scheme,
a free relativistic massive fermion is described as an excitation
from a degree of freedom of the free Dirac quantum field. For the
free Dirac field, the energy quanta of the infinite harmonic
oscillators are the relativistic energies $E_{\vec p}$ of free
fermions or antifermions.

Finally, the Feynman propagator $\mathcal{S}_{\alpha\beta}^F(x-y)$
of the free Dirac field is defined as
$i\mathcal{S}_{\alpha\beta}^F(x-y)=\bra{0}{\hat
T}\corc{\hat\psi_{\alpha}(x)\hat{\bar\psi}_{\beta}(y)}\ket{0}$. By
using the plane wave expansion, this propagator can be re-written
as $ i\mathcal{S}_{\alpha\beta}^F(x-y)=\Dirac
\delta_{\alpha\beta}i\Delta_F(x-y)$, where $i\Delta_F(x-y)$
represents the Feynman propagator of a scalar field.


\section{Dirac oscillator in the (1+1) dimensional case}\label{sec:OscDir(1+1)}

One of the most important and well studied systems in
non-relativistic quantum mechanics is the harmonic oscillator.
This system is characterized by its simplicity and usefulness. The
harmonic oscillator is a simple system due to the fact that its
Hamiltonian operator is quadratic both in coordinates and momenta.
An analogous system in relativistic quantum mechanics is given by
a linear interaction term introduced in the Dirac equation. In
fact the Dirac equation corresponds to the linearization of
relativistic Schr\"odinger equation. So the harmonic potential
must be introduced in the Dirac equation as the quadratic root of
the quadratic potential, i. e. the potential must be linear. The
interacting system constituted by a relativistic massive fermion
under the action of a linear potential is known as the Dirac
oscillator \cite{MS89}-\cite{MS93}.

In this section we will consider the Dirac oscillator in the (1+1)
dimensional case. Initially we will obtain the energy spectrum and
the wave functions describing the quantum states of this system.
Next we develop the canonical quantization procedure for the Dirac
oscillator field in 1+1 dimensions.


\subsection{Spectrum and wave functions in (1+1) dimensions}

The linear interaction over a relativistic fermion of mass $m$
that moves in one-dimension over the $z$ axis is introduced into
the Dirac equation \cite{MS89,JB90,RM95} by substituting the
momentum $p_z$ as $\hat p_{z}\rightarrow\hat
p_{z}-im\omega\gamma^0\hat z$, where $\omega$ is the frequency
associated to the oscillator. Thus, the one-dimensional Dirac
oscillator equation takes the form ($\hbar=c=1$)
\begin{align}\label{eq:DO}
i\parc{}{t}\ket\psi=\corc{\alpha_{3}\cdot\corc{\hat p_{z}-im\omega
\beta\hat z}+\beta m}\ket\psi.
\end{align}
The solution of equation (\ref{eq:DO}) allows to obtain both the
energy spectrum and the wave functions describing the quantum
states of this system. As we will show furtherly, these wave
functions will be used in the canonical quantization procedure of
this system. Given that the interaction does not mix positive and
negative energies, it can be possible to rewrite the
four-component spinor state in eq. (\ref{eq:DO}) as
\begin{align}\label{eq:BE}
\ket\psi=\begin{pmatrix}
              \ket{\phi}\\
              \ket{\chi}
            \end{pmatrix},
\end{align}
where $\ket{\phi}$ and $\ket{\chi}$ are spinors. If we use the
standard representation of the Dirac matrices, the time
independent Hamiltonian equation takes the form

{\scriptsize
\begin{align}
  \begin{pmatrix}
    m & \sigma_{3}\cdot\corc{\hat p_{z}+im\omega\hat z} \\
    \sigma_{3}\cdot\corc{\hat p_{z}-im\omega\hat z} & -m \\
  \end{pmatrix}
  \begin{pmatrix}
         \ket{\phi}\\
         \ket{\chi}
  \end{pmatrix}=E
  \begin{pmatrix}
         \ket{\phi}\\
         \ket{\chi},
  \end{pmatrix}
\end{align}}
which leads to the following coupled equations
\begin{subequations}
\begin{align}
(E-m)\ket{\phi}=\sigma_{3}\cdot\corc{\hat p_{z}+im\omega\hat
z}\ket{\chi},\label{eq:Fi1}\\
(E+m)\ket{\chi}=\sigma_{3}\cdot\corc{\hat p_{z}-im\omega\hat
z}\ket{\phi}.\label{eq:Fi2}
\end{align}
\end{subequations}
Starting from (\ref{eq:Fi1}) and (\ref{eq:Fi2}), we obtain that
\begin{align}\label{eq:OD2}
\corc{\hat p_{z}^{2}+(m\omega)^{2}\hat z^{2}}\ket{\psi}=
\corc{(E^{2}-m^{2})\mathbb{I}+ m\omega\beta}\ket{\psi},
\end{align}
where the commutation relations between the operators $\hat p_z$
and $\hat z$ have been taken into account. In the last equation,
$\mathbb{I}$ is the identity matrix $4\times 4$ and $\beta$ is the
Dirac matrix $\gamma^0$. Using the following reparametrization
\cite{JB90}
\begin{subequations}
\begin{align}
\hat \zeta&=(m\omega)^{\frac{1}{2}}\hat z,\\
\hat p_{\zeta}&=(m\omega)^{-\frac{1}{2}}\hat p_{z},
\end{align}
\end{subequations}
it is possible to write the last equation as
\begin{align}\label{eq:OD3}
\corc{\hat
p_{\zeta}^{2}+\hat\zeta^{2}}\ket{\psi}=\eta_{\pm}\ket{\psi},
\end{align}
where
\begin{align}
\eta_{\pm}=\frac{E^{2}-m^{2}}{m\omega}\pm 1.
\end{align}
We note that the equations (\ref{eq:Fi1}) and (\ref{eq:Fi2}), for
the states $\ket{\phi}$ and $\ket{\chi}$, allow to the equation
(\ref{eq:OD3}) that has the form of a harmonic oscillator
equation. This result suggests us that we can introduce the
creation and annihilation operators given by
\begin{align}
\hat a^{\dag}=\frac{1}{\sqrt{2}}\corc{\hat\zeta-i\hat
p_{\zeta}},\quad \hat a=\frac{1}{\sqrt{2}}\corc{\hat\zeta+i\hat
p_{\zeta}},
\end{align}
with the purpose to obtain the eigenvalues of the system. In this
way, using these operators, the equations (\ref{eq:Fi1}) and
(\ref{eq:Fi2}) have the form
\begin{subequations}
\begin{align}
\ket\phi&=i\frac{\sqrt{2m\omega}}{E-m}\sigma_3\crea{a}{}\ket\chi,\label{eq:Fi3}\\
\ket\chi&=-i\frac{\sqrt{2m\omega}}{E+m}\sigma_3\des{a}{}\ket\phi.\label{eq:Fi4}
\end{align}
\end{subequations}
Substituting again (\ref{eq:Fi3}) into (\ref{eq:Fi4}), it is
possible to find the following equation for the state $\ket\phi$
\begin{align}
\ket\phi&=\frac{2m\omega}{E_n^2-m^2}\hat{N}\ket\phi,\label{eq:Est1}
\end{align}
while the state $\ket\chi$ satisfies
\begin{align}
\ket\chi&=\frac{2m\omega}{E_{n'+1}^2-m^2}(\hat{N}+1)\ket\chi,\label{eq:Est2}
\end{align}
where $\hat N=\hat a^{\dag}\hat a$ is the occupation number
operator of the state. Since $\hat N$ determines the energy level
of the state where the particle (antiparticle) is, then the energy
spectrum can be deduced from (\ref{eq:Est1}) and (\ref{eq:Est2}).
Therefore we have
\begin{itemize}
\item From (\ref{eq:Fi3}), the energy spectrum for positive energy states is
\begin{align}
E_{n}^2=2nm\omega+m^{2}.
\end{align}
\item From (\ref{eq:Fi4}), the energy spectrum for negative energy states is
\begin{align}
E_{n'+1}^2=2(n'+1)m\omega+m^{2}.
\end{align}
\end{itemize}
These energy spectrums for fermions and antifermions can be
written simultaneously if we impose the following quantum number
condition
\begin{align}
n'=n-1,\qquad\qquad\text{with}\ n\ne0,
\end{align}
thus the total spectrum can be written as \cite{RS01}
\begin{align}
E_{n}=\pm\corc{2\abs nm\omega+m^{2}}^{\frac{1}{2}},
\qquad\text{con}\ n\in \mathbb{Z}.\label{eq:TEE}
\end{align}
The upper sign in this expression is taken for $n\ge 0$, while the
lower sign is for $n<0$ \cite{RS01}, therefore $E_{-n}=-E_n\
(\text{for}\ n\ne 0)$, i. e. the negative quantum numbers
correspond to the negative energy states. We observe from
(\ref{eq:TEE}), that the lower positive energy state whose energy
value is $m$ corresponds to the state with $n=0$, while the
greater negative energy state whose energy value is $-\corc{2
m\omega+m^{2}}^{\frac{1}{2}}$ corresponds to the state $n=-1$.
Additionally, we observe that if $\omega \ll m$, then the energy
difference between the states $\phi_0$ and $\chi_{-1}$ is $\Delta
E = 2m + \omega$. For $\omega=0$, i. e. the harmonic potential
vanishes into eq. (\ref{eq:DO}), then $\Delta E = 2m$, which is a
well known result obtained from the Dirac equation in the free
case.

\begin{figure*}[htb]
  \centering
  \setlength{\unitlength}{1pt}
  \begin{picture}(200,200)
    \thicklines\put(20,190){E}\put(0,120){m}\put(-5,80){-m}
    \put(10,0){\vector(0,1){200}} \put(0,100){\vector(1,0){200}}
    \put(10,120){\line(1,0){160}}\put(180,120){$E_0$}
    \put(10,140){\line(1,0){160}}\put(180,140){$E_1$}
    \put(10,160){\line(1,0){160}}\put(180,160){$E_2$}
    \put(10,180){\line(1,0){160}}\put(180,180){$E_3$}
    \put(10,60){\line(1,0){160}}\put(180,60){$E_{-1}$}
    \put(10,40){\line(1,0){160}}\put(180,40){$E_{-2}$}
    \put(10,20){\line(1,0){160}}\put(180,20){$E_{-3}$}
    \put(60,20){\vector(0,1){20}}\put(70,30){$\crea{a}{}$}
    \put(120,40){\vector(0,-1){20}}\put(130,30){$\des{a}{}$}
    \put(60,160){\vector(0,1){20}}\put(70,170){$\crea{a}{}$}
    \put(120,180){\vector(0,-1){20}}\put(130,170){$\des{a}{}$}
    \put(60,60){\vector(0,1){60}}\put(70,80){$\crea{a}{}\ket{\psi_{-1}}$}
    \put(120,120){\vector(0,-1){60}}\put(130,80){$\des{a}{}\ket{\psi_0}$}
  \end{picture}
  \caption{Energy spectrum of the one-dimensional Dirac oscillator and
  the action of the creation and annihilation operator on some system states.}
  \label{fig:EspEn}
\end{figure*}
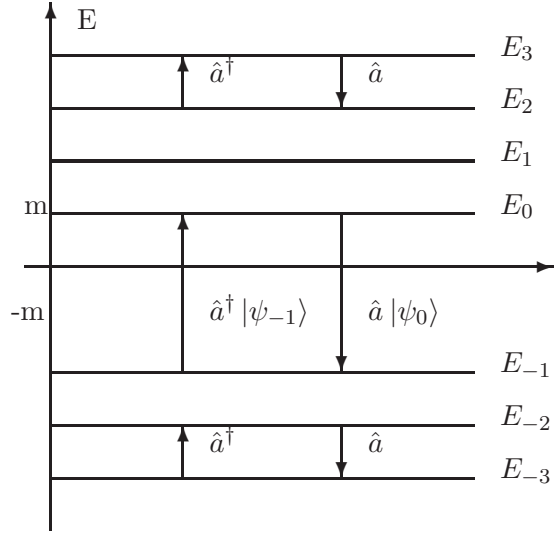

By using the previous results, the states of the system can be
written as
\begin{align}
\ket{\psi_n}= \begin{pmatrix}
                \ket{\phi_n}\\
                \ket{\chi_n}
               \end{pmatrix},
\end{align}
where the quantum number $n$, which is an integer number, can
describe positive and negative energy states. Using the expression
(\ref{eq:Fi4}), we can write that
\begin{align}
\ket{\psi_n}= \begin{pmatrix}
                \ket{\phi_n}\\
                -i\frac{\sqrt{2m\omega}}{E+m}\sigma_3\des{a}{}\ket{\phi_n}
               \end{pmatrix}.
\end{align}
If we apply the occupation number operator $\hat N$ over the state
of a system described by $\ket{\psi_n}$, we obtain
\begin{align}
 \hat N\ket{\psi_n}=\begin{pmatrix}
                \abs{n}\ket{\phi_n}\\
                (\abs{n}-1)\ket{\chi_n}
               \end{pmatrix},
\end{align}
where we have assumed that the state $\ket{\phi_n}$ has an
occupation number given by $\abs{n}$ and where we have used the
relation (\ref{eq:Fi4}) and the properties of the creation and
annihilation operators. For the last expression, we realize that
the lowest spinor $\ket{\chi_n}$ has associated the occupation
number given by $\abs{n}-1$. Thus the states $\ket{\phi_n}$ and
$\ket{\chi_n}$ can be written as
\begin{subequations}\label{eq:R}
\begin{align}
 \ket{\phi_n}&=\ket n\xi_n^1,\\
 \ket{\chi_n}&=\ket{n-1}\xi_n^2,
\end{align}
\end{subequations}
where $\xi_n^{1,2}$ represent the two-component spinors and
$\ket{n}$ represents a state with occupation number $\abs{n}$. In
consequence, the states of the system are rewritten as
\begin{align}
\ket{\psi_n}= \begin{pmatrix}
                \ket n\xi_n^1\\
                \ket{n-1}\xi_n^2
               \end{pmatrix}.
\end{align}
Taking into account that the creation $\crea{a}{}$ and
annihilation $\des{a}{}$ operators satisfy that
$\crea{a}{}\ket{n}=\sqrt{\abs n+1}\ \ket{n+1}$,
$\des{a}{}\ket{n}=\sqrt{\abs n}\ \ket{n-1}$, then these operators
acting on the state $\ket{\psi_n}$ allow to
\begin{subequations}
\begin{align}
 \crea{a}{}\ket{\psi_n}&=\begin{pmatrix}
                          \sqrt{\abs n+1}\,\ket{n+1}\xi_{n+1}^1\\
                          \sqrt{\abs n}\,\ket{n}\xi_{n+1}^2
                         \end{pmatrix}
,\ \text{for}\ n\ne-1,\\
 \des{a}{}\ket{\psi_n}&=\begin{pmatrix}
                          \sqrt{\abs n}\,\ket{n-1}\xi_{n-1}^1\\
                          \sqrt{\abs{n-1}}\,\ket{n-2}\xi_{n-1}^2
                         \end{pmatrix}
,\ \text{for}\ n\ne0,
\end{align}
\end{subequations}
whereas these operators acting on the states $\ket{\psi_0}\
\text{y}\ \ket{\psi_{-1}}$, which have associate the occupation
numbers $n=0,-1$, respectively, allow to
\begin{subequations}
\begin{align}
\des{a}{}\ket{\psi_0}&=\frac{1}{2}\corc{1-\beta}\ket{\psi_{-1}},\\
\crea{a}{}\ket{\psi_{-1}}&=\sqrt{2}\ket{\psi_0}.
\end{align}
\end{subequations}
In figure \ref{fig:EspEn} we have schematically represented the
action of the creation and annihilation operators on the positive
and negative energy states. We observe that the effect of the
annihilation operator on the state $\ket{\psi_0}$, which
corresponds to the lowest positive energy state, is such that it
does not annihilate the state but it drives it to the state
$\ket{\psi_{-1}}$, which corresponds to the greater negative
energy state. Likewise, the effect of applying the creation
operator on the state $\ket{\psi_{-1}}$ is such that it does not
annihilate that state, but drives it to the state $\ket{\psi_0}$.
Therefore, we observe that the appearing of the negative energy
states generates the well known problem of the Dirac theory: a
minimal energy state does not exist, then it is possible to obtain
an infinite energy amount from this system. In order to give a
solution to this problem, it is necessary to introduce the Dirac's
sea picture for the Dirac oscillator which will be
performed by means of the canonical quantization for this system. \\

After calculating the energy spectrum of the one-dimensional Dirac
oscillator, we proceed to obtain the wave functions. We substitute
(\ref{eq:BE}) into (\ref{eq:OD3}), then we obtain the following
differential equation for the wave function associated to the
bispinor $\ket{\phi}$
\begin{align}\label{eq:OD4}
\esp{\dder{}{\zeta}+\corc{\eta_+-\zeta^{2}}}\phi(\zeta)=0,
\end{align}
where we have used the coordinate representation of the wave
function given by $\phi(\zeta)=\bra{\zeta}\ket{\phi}$. The
differential equation (\ref{eq:OD4}) corresponds to the one of a
relativistic harmonic oscillator, whose solution is \cite{RS01}
\begin{align}
\phi_n(\zeta)=N_{\abs n}H_{\abs
n}(\zeta)e^{-\frac{\zeta^2}{2}}\xi_n^1.
\end{align}
Likewise the solution to the differential equation associated to
the wave function $\chi_n(\zeta)=\bra{\zeta}\ket{\chi_n}$ has the
form
\begin{align}
\chi_n(\zeta)=N_{\abs n-1}H_{\abs
n-1}(\zeta)e^{-\frac{\zeta^2}{2}}\xi_n^2.
\end{align}
In the expressions for $\phi_n(\zeta)$ and $\chi_n(\zeta)$,
$H_{n}(\zeta)$ represent the Hermite polynomials. Now, from the
expression \eqref{eq:Fi4} we can obtain the following relation
between the spinors $\xi_n^1$ and $\xi_n^2$
\begin{align}
\xi_n^2=-i\sqrt\frac{E_n-m}{E_n+m}\sigma_3\ \xi_n^1,
\end{align}
where we have used the properties of the creation and annihilation
operators and the definitions given by \eqref{eq:R}. Now it is
possible to write
\begin{subequations}
\begin{align}
\xi^1_n&=\begin{pmatrix}
             \sqrt{\frac{E_n+m}{2E_n}}\\
             0
            \end{pmatrix},\\
\xi^2_n&=\begin{pmatrix}
             -i\sqrt{\frac{E_n-m}{2E_n}}\\
             0
            \end{pmatrix}.
\end{align}
\end{subequations}
We observe that the spinor $\xi^2_n$ is annihilated for the case
$n=0$, which implies $\ket{\chi_0}\equiv 0$. So the most general
solution for the one-dimensional Dirac oscillator equation
(\ref{eq:DO}) is given by \cite{RS01,RR07}
\begin{align}\label{eq:SOD}
\psi_{n}(z,t)=&\sqrt{m\omega}\begin{pmatrix}
                    &\phi_n(z)\xi^1_n\\
                    &\chi_n(z)\xi^2_n
                  \end{pmatrix}e^{-iE_nt},
\end{align}
where the normalization for the wave functions has been performed
and $n=0,\pm1,\pm2,\ldots$. Finally, we can write the
one-dimensional Dirac oscillator equation (\ref{eq:DO}) in its
explicit covariant form \cite{JB90,RM95}
\begin{align}\label{eq:EODC}
\DiracOsc\psi=0,
\end{align}
where $\sigma^{\mu\nu}=\tfrac{i}{2}[\gamma^{\mu},\gamma^{\nu}]$
and the field-strength tensor $F_{\mu\nu}$ associated to the
harmonic interaction presents in this system is given by
\begin{align}
F_{\mu\nu}=\partial_{\mu}A_{\nu}-\partial_{\nu}A_{\mu},
\end{align}
where we have defined the fourpotential associated to the
interaction as
\begin{align}
A_{\mu}=\frac{1}{4}\corc{2(u\cdot x)x_{\mu}-x^{2}u_{\mu}}.
\end{align}
In the last expression $u_{\mu}=(m\omega,\vec{0})$ is a fourvector
that depends on the reference frame. In the references
\cite{JB90,RM95}, two different physical pictures for the
interaction of the Dirac oscillator have been considered.

\subsection{Canonical quantization procedure in (1+1) dimensions}
By means of the previously defined quantities, we introduce the
Lagrangian density of the system as
\begin{align}
\mathcal{L}=\bar\psi\Dirac\psi+\bar\psi\sigma^{\mu\nu}\psi\,F_{\mu\nu},
\end{align}
which, by using the Euler-Lagrange motion equation, gives the
one-dimensional Dirac oscillator equation (\ref{eq:EODC}). As the
Lagrangian density has an explicit dependence on the position
coordinate $x$, it is not possible to determine the
energy-momentum tensor of the system \cite{RH96}. However, it can
be found that the Hamiltonian density is
\begin{align}
\mathcal{H}=\psi^{\dag}(-i\alpha_{3}\cdot(\partial_{z}+m\omega\beta
z)+ \beta m)\psi.
\end{align}
To follow the standard canonical quantization procedure
\cite{GR96}, we initially impose the following commutation
relations of the Jordan-Wigner type for the fermion fields
\begin{align}
\llav{\hat\psi_{\alpha}(z,t),\hat\psi^{\dag}_{\beta}(z',t)}&=
\delta_{\alpha\beta}\delta(z-z'),\label{eq:RC1}\\
\llav{\hat\psi_{\alpha}(z,t),\hat\psi_{\beta}(z',t)}&=0,\label{eq:RC2}\\
\llav{\hat\psi^{\dag}_{\alpha}(z,t),\hat\psi^{\dag}_{\beta}(z',t)}&=
0\label{eq:RC3}.
\end{align}
Using these relations, we obtain the Hamiltonian operator for the
Dirac oscillator field operator $\hat \psi$ in the form
\begin{align}
\hat H=\int dz\,\hat\psi^{\dag}(-i\alpha_3\cdot(\partial_{z}
+m\omega\beta z)+\beta m)\hat\psi.
\end{align}
Considering that the Heisenberg equation for the field operator
$\hat \psi$ is given by $\dot{\hat\psi}(z,t)=-i[\hat\psi(z,t),\hat
H]$, then we can obtain that the motion equation for the Dirac
oscillator field operator is
\begin{align}
i\dot{\hat\psi}(z,t)&=\corc{-i\alpha_3\cdot\corc{\partial_{z}+m\omega\beta
z}+\beta m}\hat\psi(z,t).
\end{align}
Now we write the Dirac oscillator field operator using the wave
functions of the Dirac oscillator (\ref{eq:SOD}) as the base of
the expansion. These wave functions are written in terms of the
Hermite polynomials which represent a complete set of orthonormal
polynomials. Thus, the Fourier serie expansion for the Dirac
oscillator field operator can be written as
\begin{align}\label{eq:EFOD}
\hat\psi(z,t)&=\sum_{n=-\infty}^{\infty}\des{b}{n}\psi_n(z,t)
e^{-iE_{n}t} \nonumber \\
&=\sum_{n=0}^{\infty}\des{b}{n}u_n(z)e^{-iE_{n}t}+
\sum_{n=1}^{\infty}\des{b}{-n}\nu_{n}(z)e^{iE_{n}t} \nonumber\\
&=\hat\psi_{+}(z,t)+\hat\psi_{-}(z,t),
\end{align}
where the positive and negative energy contributions have been
separated $\hat\psi_{\pm}(z,t)$ and the spinors $u_{n}(z)$ and
$\nu_{n}(z)$ have been defined as
\begin{align}
u_{n}(z)=\begin{pmatrix}
       \phi_n(z)\xi^1_n\\
       \chi_n(z)\xi^2_n
      \end{pmatrix},\
\nu_{n}(z)=\begin{pmatrix}
       \phi_{-n}(z)\xi^1_{-n}\\
       \chi_{-n}(z)\xi^1_{-n}
      \end{pmatrix}.
\end{align}
If we make use of the anticommutation relations for the fields
(\ref{eq:RC1}), (\ref{eq:RC2}) and (\ref{eq:RC3}), we can verify
that creation $\hat{b}_{n}^{\dag}$ and annihilation $\hat{b}_{n}$
operators of positive (for $n \geq 0$) and negative (for $n < 0$)
energy particles satisfy the following anticommutation relations
\begin{align}
\llav{\hat{b}_{n},\hat{b}_{m}^{\dag}}&=\delta_{n,m},\label{eq:RC21}\\
\llav{\hat{b}_{n},\hat{b}_{m}}&=0,\label{eq:RC22}\\
\llav{\hat{b}_{n}^{\dag},\hat{b}_{m}^{\dag}}&=0\label{eq:RC23}.
\end{align}
Using these anticommutation relations, we can find that the
Hamiltonian operator is now given by
\begin{align} \label{eq:OHOD}
\hat H=&\sum_{n=-\infty}^{\infty}E_{n}\hat{b}_{n}^{\dag}\hat{b}_{n}
\nonumber\\
=&\sum_{n=0}^{\infty}E_{n}\hat{b}_{n}^{\dag}\hat{b}_{n}-
\sum_{n=1}^{\infty}E_{n}\hat{b}_{-n}^{\dag}\hat{b}_{-n},
\end{align}
where we have also used the expansion of the field operator
(\ref{eq:EFOD}) and the properties of the Hermite polynomials. We
can observe that the eigenvalues of this Hamiltonian operator can
take negative values without restriction at all, because it is
possible the creation of negative energy particles. To solve this
problem we take the Dirac's sea picture for this system, so we
impose that all the negative energy states are occupied by
negative energy particles. This configuration is called the vacuum
state of the Dirac oscillator field $\ket{0}$ that is written as
\cite{CS07}
\begin{align}\label{eq:EstVac}
 \ket{0}=\prod_{n=1}^\infty\crea{b}{-n}\ket{0_D},
\end{align}
where $\ket{0_D}$ is the Dirac vacuum state which is characterized
for not been occupied by fermions of positive or negative energy.
We observe that if we apply the creation operator of negative
energy particle $\crea{b}{-m}\ket{0}$ on the vacuum state
$\ket{0}$ we obtain
\begin{align}
 \crea{b}{-m}\ket{0}=\prod_{n=1}^\infty\crea{b}{-m}\crea{b}{-n}\ket{0_D}=0,
\end{align}
which implies that it is not possible to create a new negative
energy fermion because all the negative energy states are
occupied. On the other hand, it is possible that a negative energy
state can be annihilated and then a hole can be originated in the
Dirac's sea. This hole represents an antiparticle with negative
energy. If we use the anticommutation relations (\ref{eq:RC21}),
(\ref{eq:RC22}) and (\ref{eq:RC23}) into (\ref{eq:OHOD}), we find
that the Hamiltonian operator takes the form
\begin{align}
\hat H&=\sum_{n=0}^{\infty}E_n\crea{b}{n}\des{b}{n}+
\sum_{n=1}^{\infty}E_{n}\corc{\des{b}{-n}\crea{b}{-n}-1}.
\end{align}
If the divergent energy of the vacuum $\sum_{n=1}^{\infty}E_{n}$
is subtracted, the Hamiltonian operator can be reduced to
\begin{align}
\hat H&=\sum_{n=0}^{\infty}E_n\crea{b}{n}\des{b}{n}+
\sum_{n=1}^{\infty}E_n\des{b}{-n}\crea{b}{-n}.
\end{align}
We note that now this Hamiltonian operator does not have the
problem observed before because it has been positively defined.
Finally, we perform the following canonical transformations to
clearly differentiate between the operators associated to
particles and antiparticles \cite{GR96}
\begin{subequations}
\begin{align}
\crea{b}{n}&=\crea{c}{n},\\
\des{b}{n}&=\des{c}{n},\\
\des{b}{-n}&=\crea{d}{n},\\
\crea{b}{-n}&=\des{d}{n},
\end{align}
\end{subequations}
where it is possible to identify $\des{d}{n}$ and $\crea{d}{n}$ as
the annihilation and the creation operators of antiparticles,
respectively. These operators satisfy the anticommutation relation
$\llav{\hat{d}_{n},\hat{d}_{m}^{\dag}}=\delta_{n,m}$. These
transformations imply that now the application of the annihilation
operator of negative energy particle on the vacuum state can be
understood as the application of a creation operator of
antiparticle. Using this notation, the Fourier serie expansion for
the Dirac oscillator field (\ref{eq:EFOD}) has now the form
\begin{align}\label{eq:FEDOF}
\hat\psi(z,t)&=\sum_{n=0}^{\infty}\des{c}{n}u_n(z)e^{-iE_{n}t}
+\sum_{n=1}^{\infty}\crea{d}{n}v_{n}(z)e^{iE_{n}t}.
\end{align}
Now the Hamiltonian operator for the Dirac oscillator field can be
written as
\begin{align} \label{eq:HODOF}
\hat H&=\sum_{n=0}^{\infty}E_n\crea{c}{n}\des{c}{n}+
\sum_{n=1}^{\infty}E_n\crea{d}{n}\des{d}{n}=\sum_{n=0}^{\infty}E_n
{\hat n}^{(c)}_n+ \sum_{n=1}^{\infty}E_n{\hat n}^{(d)}_n,
\end{align}
where ${\hat n}^{(c)}_n$ represents the number operator for
fermions and ${\hat n}^{(d)}_n$ represents the number operator for
antifermions. It is possible to observe that $\crea{c}{n}$ acting
on $\ket{0}$ creates a fermion having energy $E_{n}$ and charge
$+e$, whereas $\crea{d}{n}$ creates antifermion having energy
$E_{n}$ and charge $-e$. In this scheme, an interacting
relativistic massive fermion is described as an excitation from a
degree of freedom of the Dirac oscillator quantum field. For the
Dirac oscillator field the energy quanta $E_{n}$ are the energies
of linear harmonic oscillators. We observe that, in this sense,
the Hamiltonian operator (\ref{eq:HODOF}) is analogous to the
Hamiltonian operator for the free Dirac field in three dimensions
given by ($\ref{eq:HOFDF}$).

Using the Fourier expansion (\ref{eq:FEDOF}) it is possible to
determine other relevant physical quantities. For instance, the
charge operator $\hat Q=e\int dz\,\hat\psi^{\dag}\hat\psi$
\cite{GR96} can be written as
\begin{align}
\hat Q&=e\sum_{n=0}^{\infty} \corc{\hat{b}_{n}^{\dag}\hat{b}_{n}-
\hat{c}_{n}^{\dag}\hat{c}_{n}},
\end{align}
where the non-observable charge of the vacuum has been removed.
Likewise, the momentum operator $\hat P=-i\int
dz\,\hat\psi^{\dag}\nabla\hat\psi$ \cite{GR96} can be also written
in an explicit form using the expansion of the Dirac oscillator
field.

The Feynman propagator for the Dirac oscillator field in the
coordinate space is defined as \cite{GR96,GR03}
\begin{align} \label{eq:FPCE}
i\mathcal{S}_{\alpha\beta}^F(z-z',t-t')=\bra{0}\hat
T\corc{\hat\psi_{\alpha}(z,t)\hat{\bar\psi}_{\beta}(z',t')}\ket{0}.
\end{align}
Substituting (\ref{eq:EFOD}) into (\ref{eq:FPCE}) and taking into
account the definition of the time-ordered operator $\hat T$, we
obtain {\small
\begin{align}
i\mathcal{S}_{\alpha\beta}^F(z-z',t-t')=&\,\Theta(t-t')
\sum_{n=0}^{\infty}u_{\alpha,n}(z)\bar
u_{\beta,n}(z')e^{-iE_n(t-t')}
\notag \\
&-\Theta(t'-t)\sum_{n=1}^{\infty}\nu_{\alpha,n}(z)\bar
\nu_{\beta,n}(z')e^{iE_n(t-t')}.
\end{align}}
Now we will obtain the Feynman propagator in the momenta space. To
do it, we consider firstly that the Fourier transformation for the
Hermite polynomials can be written as
\begin{align}
\mathfrak{F}\llav{e^{-\frac{x^2}{2}}H_n(x)}=(-i)^ne^{-\frac{k^2}{2}}H_n(k),
\end{align}
and secondly, that the contour integral for the energy eigenvalues
is written as \cite{RR07}
\begin{align}
 i\oint_C\frac{dp_o}{2\pi}\,\frac{e^{-ip_o(t-t')}}{p_0^2-p_n^2}
 =\,\Theta(t-t')\frac{e^{-iE_n(t-t')}}{2E_n}
 +\Theta(t'-t)\frac{e^{iE_n(t-t')}}{2E_n},
\end{align}
where $p_{n}^2=2\abs nm\omega+m^{2}$ and $E_{n}$ is given by
(\ref{eq:TEE}). Then we find that
\begin{align}
\mathcal{S}_{\alpha\beta}^F(z-z',t-t')=\oint_C\frac{dp_o\,d p_z\,d
p_{z'}}{(2\pi)^3}
\mathcal{S}_{\alpha\beta}^F(p_o,p_z,p_{z'})e^{-i\corc{p_{z}z+p_{z'}z'}},
\end{align}
where the Feynman propagator in the momenta space for the
one-dimensional Dirac oscillator field is given by
\begin{align}
 \mathcal{S}_{\alpha\beta}^F(p_o,p_z,p_{z'})=\sum_{n=0}^\infty
 \frac{(-1)^n}{p_0^2-p_n^2}\left[u_{\alpha,n}(p_z)\bar u_{\beta,n}(p_{z'})-
 v_{\alpha,n-1}(p_z)\bar v_{\beta,n-1}(p_{z'})\right].
\end{align}
This result is in agreement with the presented in \cite{RR07},
where the Feynman propagator of the one-dimensional Dirac
oscillator was obtained using the path integral formalism and
working with a different representation of the Dirac matrices.\

\section{Dirac oscillator in the (3+1) dimensional case}\label{sec:OscDir(3+1)}

In this section we will consider the Dirac oscillator in (3+1)
dimensions. Initially we will obtain the energy spectrum and the
wave functions describing the quantum states of this system. Next
we develop the canonical quantization procedure for the Dirac
oscillator field in the 3+1 dimensional case.


\subsection{Spectrum and wave functions in (3+1) dimensions}

In order to obtain the Dirac oscillator equation in the (3+1)
dimensional case, we perform the following substitution which is
analogous to the minimal substitution \cite{MS89,JB90,RM95}
\begin{align}
\hat{\mathbf{p}}\rightarrow\hat{\mathbf{p}}-im\omega\gamma^0\hat{\mathbf{r}},
\end{align}
where $\hat{\mathbf{p}}=(\hat{p}_x,\hat{p}_y,\hat{p}_z)$ and
$\hat{\mathbf{r}}=(\hat{x},\hat{y},\hat{z})$. The Dirac oscillator
equation has the form
\begin{align}\label{eq:DO3D}
i\parc{}{t}\ket\psi=\corc{\boldsymbol{\alpha}\cdot\corc{\hat{\mathbf{p}}-
im\omega\gamma^0\hat{\mathbf{r}}}+\beta m}\ket\psi.
\end{align}
The solutions of the Dirac oscillator equations are described by
the spinor states $\ket\psi$, that can be written as
\begin{align}\label{eq:Sep1}
  \ket\psi=
  \begin{pmatrix}
    \ket\varphi\\
    \ket\chi
  \end{pmatrix},
\end{align}
where $\ket\varphi$ and $\ket\chi$ are bispinors. The bispinors
obey the following equations
\begin{subequations}
  \begin{align}
    \corc{\hat{\mathbf{p}}²+(m\omega)^2\hat{\mathbf{r}}^2}\ket\varphi&=
    (E^2-m^2+(3+4\hat{\mathbf{s}}\cdot\hat{\mathbf{L}})m\omega)\ket\varphi,\\
    \corc{\hat{\mathbf{p}}²+(m\omega)^2\hat{\mathbf{r}}^2}\ket\chi&=
    (E^2-m^2-(3+4\hat{\mathbf{s}}\cdot\hat{\mathbf{L}})m\omega)\ket\chi.
  \end{align}
\end{subequations}
From these equations, it is evident that the Dirac oscillator in
the (3+1) dimensional case presents a strong spin-orbit coupling
term \cite{JB90,RM95}. It has been demonstrated that the Dirac
oscillator is a system where the angular momentum is conserved
\cite{JB90,RM95}, thus the quantum numbers of full angular
momentum $j$ and parity are good quantum numbers. Therefore, it is
convenient to separate the energy eigenfunctions in two parts:
radial and angular. The spatial coordinate representation of the
states \eqref{eq:Sep1} allows to the spinor wave functions given
by
\begin{align}\label{eq:EspSol}
  \psi_{n, \kappa, g}(\vec{r})=\frac{1}{r}
  \begin{pmatrix}
    F_{n,\kappa}(r)\mathcal{Y}_{\kappa, g}(\theta,\phi)\\
    iG_{n,\kappa}(r)\mathcal{Y}_{-\kappa, g}(\theta,\phi)
  \end{pmatrix},
\end{align}
where $n$ is the principal quantum number, $\kappa$ is a quantum
number related with the angular momentum and parity, and $g$ is a
quantum number related to the projection of the angular momentum
in the $z$ axis. The quantum number $\kappa$ is defined as
\cite{G00}
\begin{align}
  \kappa=\mp\corc{j+\frac{1}{2}}=
  \begin{cases}
    -(l+1), & \text{if} \quad j=l+\frac{1}{2},\\
    l, & \text{if} \quad j=l-\frac{1}{2}.
  \end{cases}
\end{align}
The full angular momentum $j$ takes values $j=l\pm\frac{1}{2}$,
where $l$ is the angular momentum and $\frac{1}{2}$ is the spin of
the fermion. The angular momentum $l'$ associated to the upper and
lower components is \cite{G00}
\begin{align}
  l'=2j-l=
  \begin{cases}
    l+1, & \text{if} \quad j=l+\frac{1}{2},\\
    l-1, & \text{if} \quad j=l-\frac{1}{2},
  \end{cases}
\end{align}The spinorial spherical harmonics
$\mathcal{Y}_{\kappa, g}(\theta,\phi)$ and $\mathcal{Y}_{-\kappa,
g}(\theta,\phi)$ are given by \cite{G00}
\begin{subequations}
  \begin{align}
    \mathcal{Y}_{\kappa, g}(\theta,\phi)&=
    \begin{pmatrix}
      \sqrt\frac{j+g}{2j}Y_{l,g-\frac{1}{2}}(\theta,\phi)\\
      \sqrt\frac{j-g}{2j}Y_{l,g+\frac{1}{2}}(\theta,\phi)
    \end{pmatrix},\\
    \mathcal{Y}_{-\kappa, g}(\theta,\phi)&=
    \begin{pmatrix}
      -\sqrt\frac{j-g+1}{2j+2}Y_{l',g-\frac{1}{2}}(\theta,\phi)\\
      \sqrt\frac{j+g+1}{2j+2}Y_{l',g+\frac{1}{2}}(\theta,\phi)
    \end{pmatrix}.
  \end{align}
\end{subequations}
Substituting \eqref{eq:EspSol} into the Dirac equation
\eqref{eq:DO3D} represented in coordinate space, we can find the
following coupled equation system for the radial functions
$F_{n,\kappa}(r)$ and $G_{n,\kappa}(r)$ \cite{RS01}
\begin{subequations}
  \begin{align}
    \esp{\der{}{r}+\frac{\kappa+m\omega r^2}{r}}F_{n,\kappa}(r)&=
    (E+m)G_{n,\kappa}(r),\\
    \esp{-\der{}{r}+\frac{\kappa+m\omega r^2}{r}}G_{n,\kappa}(r)&=
    (E-m)F_{n,\kappa}(r).
  \end{align}
\end{subequations}
The solutions of this equation system are \cite{JB90,RM95,RS01}
\begin{subequations}
  \begin{align}
    F_{n,\kappa}(r)&=A[\sqrt{m\omega}r]^{l+1}e^{-\frac{m\omega r^2}{2}}
    L_n^{l+\frac{1}{2}}(m\omega r^2),\\
    G_{n,\kappa}(r)&=\pm\text{sgn}(\kappa)A'[\sqrt{m\omega}r]^{l'+1}
    e^{-\frac{m\omega r^2}{2}}L_n^{l'+\frac{1}{2}}(m\omega r^2),
  \end{align}
\end{subequations}
where $L_n^{l}(x)$ are the Laguerre associated polynomials and
$A$, $A'$ are normalization constants given by
\begin{subequations}
\begin{align}
A=&\esp{\frac{\sqrt{m\omega}\abs{n}!\,(E_{n,\kappa}+m)}{\Gamma(\abs{n}+l+3/2)
E_{n,\kappa}}}^{\frac{1}{2}},\\
A'=&\esp{\frac{\sqrt{m\omega}\abs{n'}!\,(E_{n,\kappa}-m)}{\Gamma(\abs{n'}+l'+3/2)
E_{n,\kappa}}}^{\frac{1}{2}},
\end{align}
\end{subequations}
where the quantum number $\abs{n'}$ takes the following values
\begin{align}
  \abs{n'}=
  \begin{cases}
    \abs{n}-1, & \text{for}\quad \kappa<0\\
    \abs{n}, & \text{for}\quad \kappa>0
  \end{cases}.
\end{align}
The radial functions $F(r)$ and $G(r)$ obtained here have the same
radial structure as the one associated to the non-relativistic
three-dimensional harmonic oscillator. On the other hand, the
energy spectrum for this case depends on the $\kappa$ value. It is
possible obtain that the energy eigenvalues are \cite{RS01}
\begin{itemize}
\item For $\kappa<0$
  \begin{align}
    E_{n, \kappa}=&\pm\sqrt{m^2+4\abs{n}m\omega},
  \end{align}
where the quantum number $n$ can take the values $n=0,\pm 1,\pm
2,\ldots$, and the positive sign is chosen for $n\ge 0$, meanwhile
the negative sign is chosen for $n<0$.
\item For $\kappa>0$
  \begin{align}
    E_{n, \kappa}=&\pm\sqrt{m^2+4\corc{\abs{n}+l+\frac{1}{2}}m\omega},
  \end{align}
  where $n=\pm 0,\pm 1,\pm 2,\ldots$ and the sign is chosen as was mentioned
  in the before item. Moreover, the energy spectrum satisfies the following
  symmetry condition
  $E_{-n, \kappa}=-E_{n, \kappa}$, except for $n=0$ which means $\kappa<0$.
  For this case, it is necessary to differentiate between the quantum numbers
  $n=+0$ and $n=-0$ \cite{RS01}.
\end{itemize}
With the purpose of implementing the canonical quantization
procedure for the Dirac oscillator in (3+1) dimension, we observe
that the spinorial wave functions (\ref{eq:Sep1}) satisfy the
following orthonormality and completeness relations \cite{RS01}
\begin{subequations}
\begin{align}
\sum_{\substack{\kappa=-\infty\\\kappa\ne0}}^{\infty}\sum_{g=-\abs{\kappa}+
\frac{1}{2}}^{\abs{\kappa}-\frac{1}{2}}\sum_{n=-\infty}^{\infty}\psi_{n,\kappa,
g} (\vec r)\psi_{n,\kappa, g}^\dag(\vec r')&=\delta^3(\vec r-\vec
r')\mathbb{I}_4,
\label{eq:OrtoCon}\\
\int d^3r\,\psi_{n,\kappa, g}^\dag(\vec r)\psi_{n',\kappa',
g'}(\vec r)&=\delta_{n,n'}\delta_{\kappa,\kappa
'}\delta_{g,g'}\label{eq:ComCon},
\end{align}
\end{subequations}
respectively. Finally, we note that the covariant Dirac oscillator
equation for the (3+1) dimensional case has the same form as the
equation \eqref{eq:EODC} for the (1+1) dimensional case.

\subsection{Canonical quantization in (3+1) dimensions}

To proceed with the canonical quantization for the Dirac
oscillator field in the (3+1) dimensional case, we observe that
the form of the Lagrangian density for this case is similar to the
one in the (1+1) dimensional case and it is given by
\begin{align}
\mathcal{L}=\bar\psi\Dirac\psi+\bar\psi\sigma^{\mu\nu}\psi\,F_{\mu\nu}.
\end{align}
As in the (1+1) dimensional case, the Lagrangian density depends
explicitly on the spatial coordinates and then it is not possible
to obtain the energy-momentum tensor of the system. However, by
means of a Legendre transformation, we can obtain the following
Hamiltonian density
$\mathcal{H}=\psi^{\dag}(-i\boldsymbol{\alpha}\cdot(\vec{\nabla}+
m\omega\beta\vec{r})+\beta m)\psi$. Thus, the Hamiltonian of the
field is
\begin{align}
H=\int
d^3r\,\psi^{\dag}(-i\boldsymbol{\alpha}\cdot(\vec{\nabla}+m\omega\beta\vec{r})+
\beta m)\psi.
\end{align}
To perform the canonical quantization for this case, as is usual,
the fields are considered as field operators. Then we establish
the following Jordan-Wigner commutation relations
\begin{align}\label{eq:RC3d}
\llav{\hat\psi_{\alpha}(\vec{r},t),\hat\psi^{\dag}_{\beta}(\vec{r}',t)}&=
\delta_{\alpha\beta}\delta^3(\vec{r}-\vec{r}'),\\
\llav{\hat\psi_{\alpha}(\vec{r},t),\hat\psi_{\beta}(\vec{r}',t)}&=0,\\
\llav{\hat\psi^{\dag}_{\alpha}(\vec{r},t),\hat\psi^{\dag}_{\beta}(\vec{r}',t)}&=
0,
\end{align}
that allow us to write the Hamiltonian operator as
\begin{align}
\hat H=\int
d^3r\,\hat\psi^{\dag}(-i\boldsymbol{\alpha}\cdot(\vec{\nabla}+m\omega\beta\vec{r})+
\beta m)\hat\psi.
\end{align}
Now we can expand the field operator $\hat\psi(\vec{r},t)$, using
the spinor wave functions (\ref{eq:Sep1}), as follows
\begin{align}\label{eq:Exp3D}
\hat\psi(\vec{r},t)=\sum_{\substack{\kappa=-\infty\\\kappa\ne0}}^{\infty}
\sum_{g=-\abs{\kappa}+\frac{1}{2}}^{\abs{\kappa}-\frac{1}{2}}\sum_{n=-
\infty}^{\infty}\hat{b}_{n,\kappa, g}\psi_{n,\kappa, g}(\vec
r)e^{-iE_{n,\kappa}t},
\end{align}
where the operators $\hat{b}_{n,\kappa, g}$ and
$\hat{b}_{n,\kappa, g}^\dag$, respectively, annihilates and
creates fermions in a state defined by the quantum numbers $n$,
$\kappa$ and $g$. These operators obey the following
anticommutation relations
\begin{align}
\llav{\hat{b}_{n,\kappa, g},\hat{b}_{n'\kappa'
g'}^\dag}&=\delta_{n,n'}\delta_{\kappa,\kappa'}\delta_{g,g'},\\
\llav{\hat{b}_{n,\kappa, g},\hat{b}_{n'\kappa'
g'}}&=0,\\
\llav{\hat{b}_{n,\kappa, g}^\dag,\hat{b}_{n'\kappa' g'}^\dag}&=0,
\end{align}
Using the properties of the energy eigenfunctions
(\ref{eq:OrtoCon}) and (\ref{eq:ComCon}) into the Hamiltonian
operator \eqref{eq:Exp3D}, we can obtain
\begin{align}
\hat
H=\sum_{\substack{\kappa=-\infty\\\kappa\ne0}}^{\infty}\sum_{g=-\abs{\kappa}+
\frac{1}{2}}^{\abs{\kappa}-\frac{1}{2}}\sum_{n=-\infty}^{\infty}E_{n,\kappa}
\hat{b}_{n,\kappa, g}^\dag\hat{b}_{n,\kappa, g}.
\end{align}
This operator can be rewritten by splitting the positive and
negative energy contributions and taking into account the two
types of spectrum depending on the $\kappa$ value. In this way, we
obtain
\begin{align}
\hat H&=\sum_{\kappa=1}^\infty\sum_g\sum_{n=0}^\infty E_{n,\kappa}
\hat{b}_{n, \kappa, g}^\dag\hat{b}_{n, \kappa,
g}-\sum_{\kappa=1}^\infty \sum_g\sum_{n=0}^\infty
E_{n,\kappa}\hat{b}_{-n,\kappa, g}^\dag\hat{b}_{-n,
\kappa, g}\notag\\
&\quad+\sum_{\kappa=1}^\infty\sum_g\sum_{n=0}^\infty
E_{n,-\kappa}\hat{b}_{n,-\kappa, g}^\dag\hat{b}_{n,-\kappa,
g}-\sum_{\kappa=1}^\infty\sum_g\sum_{n=1}^\infty
E_{n,-\kappa}\hat{b}_{-n,-\kappa,g}^\dag\hat{b}_{-n,-\kappa, g},
\end{align}
where we have used the convention $E_{n,-\kappa}\equiv
E_{n,\kappa}$, for $\kappa<0$, and we have omitted the limits of
the sum over $g$. It is evident that this Hamiltonian is not
defined positively as happened in the (1+1) dimensional case.
Again we use the picture of the Dirac's sea in order to solve this
problem. To do it we take the vacuum state in an analogous way as
it was defined in \eqref{eq:EstVac}
\begin{align}\label{eq:EstVac3d}
 \ket{0}=\prod_{\substack{\kappa=-\infty\\\kappa\ne0}}^{\infty}\prod_g
 \prod_{n=0}^\infty\crea{b}{-n,\kappa,g}\ket{0_D}.
\end{align}
Therefore we can rewrite the Hamiltonian operator without
considering the vacuum energy in the following way
\begin{align}
\hat H'&=\sum_{\kappa=1}^\infty\sum_g\sum_{n=0}^\infty
E_{n,\kappa} \hat{b}_{n,\kappa, g}^\dag\hat{b}_{n,\kappa,
g}+\sum_{\kappa=1}^\infty \sum_g\sum_{n=0}^\infty
E_{n,\kappa}\hat{b}_{-n,\kappa, g}
\hat{b}_{-n,\kappa, g}^\dag\notag\\
&\quad+\sum_{\kappa=1}^\infty\sum_g\sum_{n=0}^\infty
E_{n,-\kappa}\hat{b}_{n,-\kappa, g}^\dag\hat{b}_{n,-\kappa,
g}\sum_{\kappa=1}^\infty\sum_g\sum_{n=1}^\infty
E_{n,-\kappa}\hat{b}_{-n,-\kappa,g}\hat{b}_{-n,-\kappa, g}^\dag.
\end{align}
We observe that this Hamiltonian operator describes two different
types of particles and antiparticles, because particles and
antiparticles have different energy spectrums depending on the
sign of the $\kappa$ value. This value depends explicitly on the
full angular momentum $j$ due to the value of $j$ is based on the
spin state of the fermion. In this way, we perform the following
canonical transformations
\begin{subequations}
\begin{align}
\hat{b}_{n,\kappa, g}^\dag&=\crea{b}{n,\kappa, g},\\
\hat{b}_{-n,\kappa, g}&=\crea{c}{n,\kappa, g},\\
\hat{b}_{n,-\kappa, g}^\dag&=\crea{d}{n,\kappa, g},\\
\hat{b}_{-n,-\kappa, g}&=\crea{f}{n,\kappa, g},
\end{align}
\end{subequations}
where $\crea{b}{n,\kappa, g}$ is the creation operator of
particles with full angular momentum $j=l-\frac{1}{2} $, i. e.
particles with $\kappa>0$ \cite{BL07}; $\crea{c}{n,\kappa, g}$ is
the creation operator of antiparticles with full angular momentum
$j=l-\frac{1}{2} $, i. e. antiparticles with $\kappa>0$
\cite{BL07}; $\crea{d}{n,\kappa, g}$ is the creation operator of
particles with full angular momentum $j=l+\frac{1}{2} $, i. e.
particles with $\kappa<0$ \cite{BL07}; $\crea{f}{n,\kappa, g}$ is
the creation operator of antiparticles with full angular momentum
$j=l+\frac{1}{2} $, i. e. antiparticles with $\kappa<0$
\cite{BL07}.

Finally, we can obtain that the field operator for this case can
be written by means of the expansion
\begin{align} \label{eq:ef3dfo}
\hat\psi(\vec{r},t)&=\sum_{\kappa=1}^\infty\sum_g\sum_{n=0}^\infty\des{b}{n,\kappa,
g} \psi_{n,\kappa,g}(\vec
r)e^{-iE_{n,\kappa}t}+\sum_{\kappa=1}^\infty\sum_g\sum_{n=0}^\infty
\crea{c}{n,\kappa, g}\psi_{-n,\kappa,g}(\vec r)e^{iE_{n,\kappa}t}\notag\\
&\quad+\sum_{\kappa=1}^\infty\sum_g\sum_{n=0}^\infty\des{d}{n,\kappa,
g}\psi_{n,-\kappa,g}(\vec
r)e^{-iE_{n,-\kappa}t}+\sum_{\kappa=1}^\infty\sum_g\sum_{n=1}^\infty\crea{f}{n,\kappa,
g}\psi_{-n,-\kappa,g}(\vec r)e^{iE_{n,-\kappa}t}.
\end{align}
In an analogous way as was showed for the (1+1) dimensional case,
starting from the expansion for the field operator
(\ref{eq:ef3dfo}), it is possible to obtain the different relevant
physical quantities associated to the Dirac oscillator field in
(3+1) dimensions.


\section{Conclusions}

In this work we have performed the canonical quantization of the
Dirac oscillator field in (1+1) and (3+1) dimensions. This
quantization has been possible because the solutions of the Dirac
oscillator equation do not present the Klein paradox \cite{DA92}.
If this fact would not have satisfy, then it had not been possible
to distinguish between positive and negative energy states
\cite{RH96}, which had restricted the possibility to perform a
Fourier expansion of the field operator. The Dirac oscillator
field has been quantized following a similar procedure as if this
field were free \cite{GR96}. However, this procedure implies
differences with respect to the Dirac free field quantization. For
instance, the Dirac oscillator is an interacting system in which
there exist bound states define with specific quantum numbers \cite{DA90}.
In this case the field operators create and annihilate fermions with
well determined energy values in contrast to what happens in the free
field case where the states have a well defined momentum \cite{AB07}.
Moreover, for this case, the momentum operator depends explicitly
on the time thus the fermions created do not have a determined
momentum. We have found that for the Dirac oscillator field it has
been impossible to obtain the energy-momentum tensor due to the
fact that the Lagrangian density has an explicit dependence on the
spatial coordinates. Nevertheless, this problem could be solved by
the introduction in the system of an additional field which
describes the interaction. We have also obtained the Feynman
propagator in the (1+1) dimensional case in agreement with the
result obtained in the literature by using an functional
procedure \cite{RR07}. We note that for the Dirac oscillator field
in the (3+1) dimensional case, we have found that there exist two
types of particles and antiparticles because there are two possible
values for the full angular momentum. Finally, we have found that
while for the free Dirac field the energy quanta of the infinite
harmonic oscillators are relativistic energies of free particles,
for the Dirac oscillator quantum field the quanta are energies of
relativistic linear harmonic oscillators. The canonical quantization
procedure for the Dirac oscillator field in (2+1) dimensions is an
exercise to develop explicitly. We consider that the possibility
to study the Dirac oscillator as a quantum field opens the doors to
future applications in different areas of the Physics.

\section*{Acknowledgments}

We thank Maurizio De Sanctis and Antonio S\'anchez for stimulating
discussions and suggestions about some aspects presented in this
paper. C. J. Quimbay thanks Vicerrector\'ia de Investigaciones of
Universidad Nacional de Colombia by the financial support received
through the research grant "Teor\'ia de Campos Cu\'anticos
aplicada a sistemas de la F\'isica de Part\'iculas, de la F\'isica
de la Materia Condensada y a la descripci\'on de propiedades del
grafeno".


\end{document}